# Phase fluctuations and the pseudogap in $YBa_2Cu_3O_x$

C. Meingast,[1] V. Pasler,[1] P. Nagel,[1] A. Rykov,[2] S. Tajima[2] and P. Olsson[3]

[1]*Forschungszentrum Karlsruhe, Institut für Festkörperphysik, 76021 Karlsruhe, Germany*
[2]*Superconductivity Research Lab-ISTEC, 10-13 Shinonome 1-chome, Koto-ku, Tokyo, Japan*
[3]*Department of Theoretical Physics, Umeå University, S90187 Umeå, Sweden*

The thermodynamics of the superconducting transition is studied as a function of doping using high-resolution expansivity data of $YBa_2Cu_3O_x$ single crystals and Monte-Carlo simulations of the anisotropic 3D-XY model. We directly show that $T_c$ of underdoped $YBa_2Cu_3O_x$ is strongly suppressed from its mean-field value ($T_c^{MF}$) by phase fluctuations of the superconducting order parameter. For overdoped $YBa_2Cu_3O_x$ fluctuation effects are greatly reduced and $T_c \approx T_c^{MF}$. We find that $T_c^{MF}$ exhibits a similar doping dependence as the pseudogap energy, naturally suggesting that the pseudogap arises from phase-incoherent Cooper pairing.

The properties of high-temperature superconductors (HTSC) vary strongly with hole doping. Especially the underdoped region of the phase diagram, where the critical temperature ($T_c$) increases with doping, has attracted considerable attention due to the gradual development of a pseudogap in the low energy electronic excitations at a temperature $T^*$, which lies significantly above $T_c$ [1-9]. Although the pseudogap phenomenon has been known for some time, the exact nature of this state and its relation to superconductivity is still a subject of great controversy [1-10]. Basically there are two competing scenarios. In the first, the opening of the pseudogap is attributed to phase incoherent precursor pairing, and $T_c$ is much smaller than $T^*$ because of strong phase fluctuations [2,8,10]. In the second scenario, the pseudogap is a normal state gap, which is independent of and competing with superconductivity [3,9]. The origin of the pseudogap is clearly of great importance for understanding the general phase diagram of HTSC.

In this Letter, we study the doping dependent thermodynamic response at $T_c$ of $YBa_2Cu_3O_x$ crystals using high-resolution dilatometry. Thermal expansion, which is closely related to the specific heat through the thermodynamic Ehrenfest or Pippard relations [11], was previously shown to be a very sensitive probe of the superconducting transition due to the large signal to background ratio and provided direct evidence of strong 3D-XY critical fluctuations at optimal doping [12]. Here we show that fluctuation effects increase dramatically and become more 2-dimensional in the underdoped region, resulting in a large depression of $T_c$ from its mean-field value ($T_c^{MF}$). We find excellent scaling of these fluctuations with Monte-Carlo specific-heat simulations of the anisotropic 3D-XY model, which clearly demonstrates the 'superconducting' origin of the fluctuations. $T_c^{MF}$ is shown to exhibit a similar doping dependence as $T^*$ naturally suggesting that the pseudogap arises from phase-incoherent Cooper pairing. Our results, thus, confirm the phase diagram based on the phase-fluctuation scenario proposed by Emery and Kivelson [10].

An untwinned YBa$_2$Cu$_3$O$_x$ single crystal [12,13] was used, whose oxygen content was varied between $x \approx 6.77$ and $x \approx 7.0$ by annealing in pure O$_2$ gas at pressures between 5 mbar (450 °C) and 376 bar (400 °C), resulting in slightly underdoped to slightly overdoped states [14]. The thermal expansivity was measured with a high-resolution capacitance dilatometer [15]. The Monte-Carlo simulations, in which anisotropy is introduced by reducing the coupling coefficient $J_z$ between spins along the z-direction, were performed using the Wolff cluster update method on systems of sizes 64x64x64 ($L_x \times L_y \times L_z$) for the isotropic case ($J_z/J_{xy}= 1$), 128x128x16 for $J_z/J_{xy} = 0.02$, and 128x128x8 for $J_z/J_{xy} = 0.004$. For the anisotropic systems the cell sizes were chosen such that $L_z/L_{xy} \approx \xi_z/\xi_{xy}$ where $\xi_z$ and $\xi_{xy}$ are the correlation lengths in the different directions and $\xi_z/\xi_{xy} = \sqrt{J_z/J_{xy}}$. Each specific heat data point is from simulations with the creation and flipping of a few million clusters.

A previous study of YBa$_2$Cu$_3$O$_x$ at optimal doping [12] showed that it is of great advantage to examine the difference between the expansivities of the b- and a-axes, $\alpha_{b-a} \equiv \alpha_b - \alpha_a$, because the anomalies at $T_c$ are of approximately equal magnitude, but of opposite sign. Taking this difference, thus, doubles the size of the anomaly and, at the same time, reduces the background [12]. In Fig. 1a $\alpha_{b-a}(T)$ is shown for 6 different values of x. Large, in comparison to the background, λ-shaped anomalies are seen at $T_c$, which decrease in size as the oxygen content is reduced from x=7.0 to x=6.77. For x=6.77, the anomalies in the a- and b-axes have the same sign and magnitude and, thus, the anomaly in $\alpha_{b-a}$ vanishes [16]. This is actually a blessing in disguise, since the $\alpha_{b-a}$ curve for x=6.77 provides us with a very good approximation of the curvature of the phonon background expansivity, $\alpha_{b-a}^{back.}$. By simply vertically scaling the smooth $\alpha_{b-a}$ (x=6.77) data until they coincide with the various data sets at high temperatures, we obtain $\alpha_{b-a}^{back.}$ for the different x values (see gray lines in Fig. 1). We believe that this procedure produces a very physical background for the different x values because the background matches the curvature of the high-temperature data very well for all x values and the variation of this background with x intuitively matches what one would expect as one progressively removes O-atoms from the chain sites. The origin of $\alpha_{b-a}^{back.}$ is the orthorhombic structure, i.e. the chains along the b-axis. The magnitude of $\alpha_{b-a}^{back.}$ is thus expected to be largest when the chains are fully oxidized (x=7.0) and to decrease as oxygen is removed from the chains (at x ≈ 6.35 the material becomes tetragonal, and $\alpha_{b-a}^{back.}$ must vanish). This is exactly the observed behavior, which gives us a great deal of confidence in $\alpha_{b-a}^{back.}$ [17].

In Fig. 1b $\alpha_{b-a}^{back.}$ has been subtracted from all original data sets yielding the electronic thermal expansivity $\alpha_{b-a}^{elec.}$. The magnitude of $\alpha_{b-a}^{elec.}$ at $T_c$ decreases drastically with oxygen depletion. This is largely due to the fact that the difference between uniaxial pressure coefficients $dT_c/dp_{b-a} \equiv dT_c/dp_b - dT_c/dp_a$ vanishes at x ≈ 6.77 [16], since $\alpha_{b-a}^{elec.}$ is directly proportional to $dT_c/dp_{b-a}$ [11]. In order to compare the shapes of these anomalies, we normalize the temperature scales by $T_c$ and then divide $\alpha_{b-a}^{elec.}(x, T/T_c)$ by $\alpha_{b-a}^{elec.}(x, T/T_c=0.7)$, which results in very good scaling in the low-temperature region ($T/T_c <0.75$, see Fig. 2a). By scaling the anomalies at low temperatures, we have in effect normalized all curves by the mean-field (MF) behavior, which allows us to quantify the effects of fluctuations relative to the MF component. For the slightly overdoped case (x=7.0), $\alpha_{b-a}^{elec.}$ resembles classical MF-behavior with small fluctuation corrections close to $T_c$. As x decreases, the magnitude of the singular part at $T_c$ increases and at the same time the range of the fluctuations increases dramatically, especially above $T_c$. For our most underdoped sample in Fig. 2a (x=6.81), the fluctuation contribution extends up to about 2 x $T_c$, and the integrated signal above $T_c$ is actually larger than the one below $T_c$. This manifests a drastic deviation from MF behavior, where all of the 'action' is below $T_c$, and clearly demonstrates the unusual character of the superconducting transition in underdoped $YBa_2Cu_3O_x$.

It is well-known that the superconducting and normal state properties of $YBa_2Cu_3O_x$ and other HTSCs are highly anisotropic and also that this anisotropy increases strongly in the underdoped region [18,19]. We now show that the unusual behavior in the underdoped region can be understood in terms of strong superconducting phase fluctuations in a system with reduced dimensionality. For this purpose we compare $\alpha_{b-a}^{elec.}$ with the specific-heat simulations of the anisotropic 3D-XY model ($C^{3DXY}$), which are shown in Fig. 2b for three different anisotropy values ($J_z/J_{xy}$=1, 0.02, and 0.004). This model belongs to the same universality class as superconductivity (in the limit $\kappa \gg 1$) and is expected to reproduce the critical behavior of the superconducting transition if one has strong fluctuations [18,20,21]. As anisotropy is introduced into the XY model, the shape of the $C^{3DXY}$ curves change in a very similar fashion as $\alpha_{b-a}^{elec.}$ curves change for decreasing x (Fig. 2a); that is, more and more of the area under the transition is shifted to higher temperatures, and the 'jump' component at $T_c$ decreases, resulting in a more symmetric anomaly. Universality implies that $C^{3DXY}$ and $C_p^{elec.}$ of $YBa_2Cu_3O_x$ follow the same scaling laws, or in other words, the anomalies should have the same shape near $T_c$ [18]. In Fig. 2b we show that this is actually the case by directly scaling our $\alpha_{b-a}^{elec.}$, which are a direct reflection of $C_p^{elec.}$, with $C^{3DXY}$. $C^{3DXY}$ for the isotropic case ($J_z/J_{xy}=1$, thick light red line in Fig. 2b) excellently matches the $\alpha_{b-a}^{elec.}$ curve near optimal doping (x=6.93, thin red line in Fig. 2b), where isotropic 3D-XY scaling has been found to be well obeyed [12,18,19]. In the underdoped region we find excellent agreement between $C^{3DXY}$ of the most anisotropic simulation ($J_z/J_{xy}=0.004$, thick light blue line) and $\alpha_{b-a}^{elec.}$ for x=6.81 (thin blue line in Fig.2b). In both cases, scaling is observed over a very wide temperature range, implying an extremely large critical region. We note that the scaling of $C^{3DXY}$ and $\alpha_{b-a}^{elec.}$ in Fig. 2b is obtained by a **single** multiplicative parameter, which can only be expected if the background subtraction in Fig. 1 is essentially correct. Also, the finite-size broadening of the transitions in both sets of data are of very similar magnitude, which greatly facilitates this comparison. In the overdoped region (x=7.0) the fluctuations decrease in size and the simulations, which are based on strong critical behavior, no longer match the experimental data.

The nearly perfect scaling of $C^{3DXY}$ and $\alpha_{b-a}^{elec.}$ around $T_c$ shown in Fig. 2b implies that the physics of the superconducting transition in under- and optimally doped YBa$_2$Cu$_3$O$_x$ is correctly accounted for by the simple anisotropic 3D-XY model, in which only phase fluctuations of the order parameter occur, since the amplitude of the spin J is fixed. For superconductors, this corresponds to a fixed amplitude of Cooper pairs, and, thus, Fig. 2b provides direct evidence for strong phase fluctuations [18,20,21] in under- and optimally doped states. We note that the large reduction in the specific heat 'jump' at $T_c$ in the underdoped region, which has been previously taken as evidence of a normal-state pseudogap [3], chain superconductivity [22], or a MF - Bose-Einstein crossover [19], is a natural consequence of the crossover from anisotropic 3D-XY behavior in the underdoped region to MF-like behavior in the overdoped region (see Fig. 2a and 2b).

It is instructive to compare the values of the anisotropy of the simulations with the closely, but not trivially, related anisotropy of the superconducting coherence lengths $\Gamma = \xi_{ab}/\xi_c$. There are two limiting cases. First, if the out-of-plane 'bare' coherence length $\xi_c^{bare}$ at $T_c$ is smaller than the spacing d between superconducting layers, then $\sqrt{J_{xy}/J_z} = \Gamma \cdot d/\xi_{ab}^{bare}$, where $\xi_{ab}^{bare}$ is the 'bare' in-plane coherence length and 'bare' refers to the fluctuation uncorrected, i.e. MF, values [21]. This appears to be the situation for x=6.81, where $\Gamma \approx 12$ [23] and $\sqrt{J_{xy}/J_z} \approx 15$ are nearly equal, which implies that $d/\xi_{a,b} \approx 1$. Such small values of the 'bare' coherence lengths at $T_c$ ($\xi_c^{bare} << 12$ Å and $\xi_{ab}^{bare} \approx 12$ Å) imply very large fluctuation corrections, since usually one expects the coherence lengths to diverge at $T_c$. The 'bare' coherence lengths, on the other hand, diverge at the fluctuation uncorrected $T_c^{MF}$, and our result suggests that $T_c^{MF}$ lies considerably higher than the real $T_c$. We will show below that this is actually the case in the underdoped region. For the second case ($\xi_c > d$), d in the above equation should be replaced by $\xi_c^{bare}$, resulting in the isotropic model ($J_z/J_{xy}=1$) [21], which we observe near optimal doping (x=6.93). Thus, although a considerable anisotropy ($\Gamma \approx 7$) still exists at optimal doping [23], the system behaves isotropic because of the strong coupling along the c-axis. Our results point to a doping-induced crossover from quasi 2D-behavior of weakly Josephson-type coupled planes in the underdoped region, where $\xi_c^{bare} < d$, to the 3D behavior of strongly coupled planes in the optimal and overdoped regions, where $\xi_c^{bare} > d$.

Phase transition temperatures are always lowered by strong fluctuations [10,18,20], and it is instructive to ask the question: how much higher would $T_c$ be if there were no fluctuations, or, in other words, what is $T_c^{MF}$? In order to extract $T_c^{MF}$ from our data, we have extrapolated the low-T MF-component of $\alpha_{b-a}^{elec.}$ beyond the real $T_c$ value, and $T_c^{MF}$ is defined as the temperature were the area under the MF curve equals the total area under $\alpha_{b-a}^{elec.}$. The inset of Fig. 3 shows this construction for x=6.88 yielding $T_c^{MF} \approx 1.25\, T_c$. This procedure, which is analogous to making an entropy conserving construction in heat capacity data, was carried out for the rest of the doping levels and the results are plotted versus doping level [24] in Fig. 3. We find the astonishing result that $T_c^{MF}$ decreases approximately linearly with increasing doping at the same time as the real $T_c$ goes over a maximum. In the overdoped region the two $T_c$ values merge because the fluctuations are greatly reduced. We note that the above estimate of $T_c^{MF}$ depends only on the shape of $\alpha_{b-a}^{elec.}$ and not on the absolute scale.

The doping dependence $T_c^{MF}$ shown in Fig. 3 presents one of our most important results, and several implications follow immediately. First, $T_c^{MF}$ and the pseudogap temperature [5], or energy scale [3,9], exhibit a very similar doping dependence, which strongly suggests that the pseudogap is due to local Cooper pairing without phase coherence. The fact that $T_c^{MF}$ also shows the same linear doping dependence as the single particle excitation gap at zero temperature $\Delta(0)$ seen in both scanning tunneling microscopy (STM) [6,7] and angle-resolved photoemission (ARPES) data [4,5], supports this conclusion, since in a MF theory $\Delta(0)$ is a direct measure of $T_c^{MF}$. The strong doping dependence of the ratio $2\Delta(0)/kT_c$ [4-7], in our view, is a fluctuation artifact. We also expect that $2\Delta(0)/kT_c$ of $Bi_2Sr_2CaCu_2O_{8+d}$ to be larger than that of $YBa_2Cu_3O_x$ at optimal doping, because $Bi_2Sr_2CaCu_2O_{8+d}$ is much more two-dimensional and fluctuations play an even more important role than in $YBa_2Cu_3O_x$ [20,25,26]. There is some evidence that this is actually the case [6,27]. Second, the maximum in $T_c$ at optimal doping is really an artifact due to fluctuations; that is, the pairing energy, as reflected by $T_c^{MF}$, decreases smoothly with increasing doping with no indication of any special behavior near optimal doping. Our phase diagram is in excellent agreement with the phase-fluctuation scenario proposed by Emery et al. [10] and also with a recent analysis of the t-J model [28]. Finally, extrapolating our $T_c^{MF}$ to even lower doping levels suggests that the strongest pairing interaction in YBa$_2$Cu$_3$O$_x$ occurs at, or near, the antiferromagnetic phase boundary. This may be an indication of a magnetic pairing mechanism, as recently proposed for several heavy-fermion compounds [29] and also suggested from magnetic neutron scattering data [30].

C.M. would like to acknowledge stimulating discussions with V. Emery, A. Junod, T. Schneider and H. Wühl. The work at SRL was supported by the New Energy and Industrial Technology Development Organization as Collaborative Research and Development of Fundamental Technologies for Superconductivity Applications.

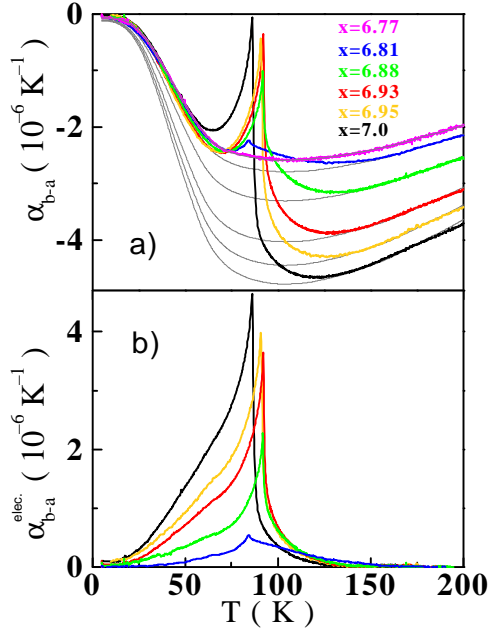

**Fig. 1.** a) Difference between the expansivity along the b- and a-axes $\alpha_{b-a}(T)$ of YBa$_2$Cu$_3$O$_x$ for x = 6.77 - 7.0. The gray lines represent the backgrounds, which were obtained by scaling the smooth x=6.77 data (see text). b) Electronic expansivity $\alpha_{b-a}^{elec.}$ obtained after subtracting the backgrounds from the original data sets.

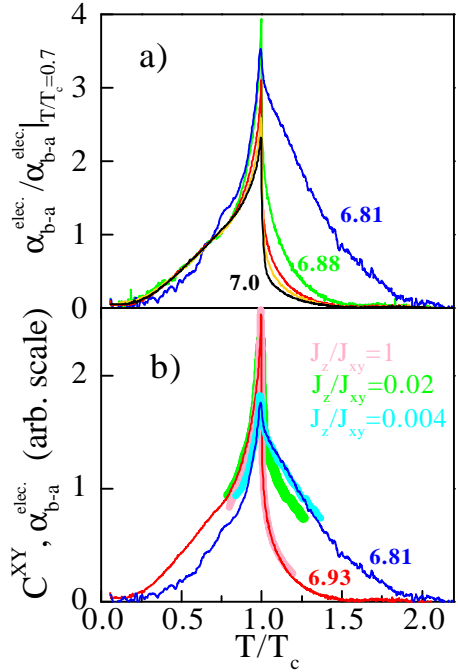

**Fig. 2.** a) Low-temperature scaled $\alpha_{b-a}^{elec.}$ showing a dramatic increase in the fluctuation signal in the underdoped samples (see text). b) Specific heat of the 3D-XY model ($C^{XY}$) for three different anisotropies (thick light-colored lines). Excellent scaling of $\alpha_{b-a}^{elec.}$ and $C^{XY}$ is observed over an extended temperature range for x=6.93 (thin red line) and x=6.81 (thin blue line).

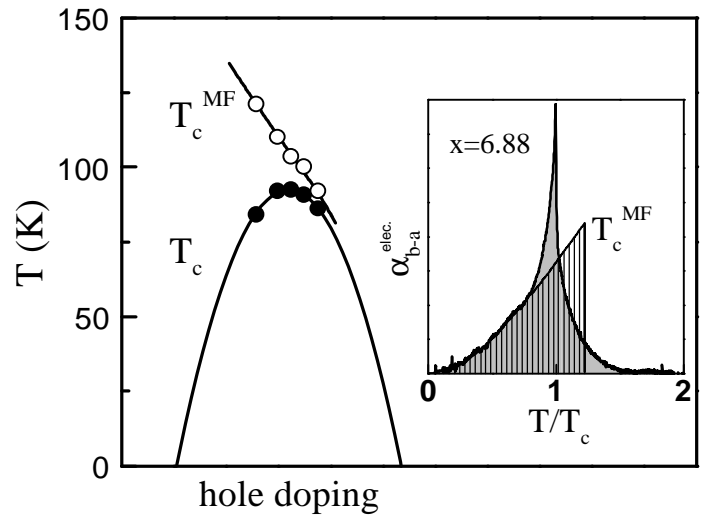

**Fig. 3**. Phase diagram of YBa$_2$Cu$_3$O$_x$ showing the doping dependence of both T$_c$ and T$_c^{MF}$. The inset shows how T$_c^{MF}$ is defined (see text).